\documentclass[a4paper]{jpconf}
\usepackage{graphicx}
\def\gsim{\mathstrut_{\displaystyle \sim}^{\displaystyle >}}
\begin{document}
\vspace*{-15mm}
\title{Non local theory of excitations applied to the Hubbard model\footnote{to be published in J. Phys.:
Conference Series (ICM'09)}}

\author{Y Kakehashi$^1$, T Nakamura$^1$, P Fulde$^2$}

\address{$^1$ University of the Ryukyus, Nishihara, Okinawa, Japan}
\address{$^2$ MPI f\"ur Physik komplexer Systeme, Dresden, Germany and 
APCTP, Pohang, Korea}

\ead{yok@sci.u-ryukyu.ac.jp}

\begin{abstract}
 We propose a nonlocal
 theory of single-particle excitations. It is based on an off-diagonal 
 effective medium and the projection operator method
 for treating the retarded Green function.  The theory determines
 the nonlocal effective medium matrix elements by requiring that they
 are consistent with those of the self-energy of the Green function.
 This arrows for a description of long-range
 intersite correlations with high resolution in momentum space.
 Numerical study for the half-filled Hubbard model on the simple cubic
 lattice demonstrates that the theory is applicable to the strong
 correlation regime as well as the intermediate regime of Coulomb
 interaction strength.  Furthermore the results show that nonlocal 
 excitations cause sub-bands in the strong Coulomb interaction regime
 due to strong antiferromagnetic correlations, decrease the
 quasi-particle peak on the Fermi level with increasing Coulomb
 interaction, and shift the critical Coulomb interaction $U_{\rm c2}$
 for the divergence of effective mass towards higher energies at least 
 by a factor of two as compared with that in the single-site approximation.
\end{abstract}

The description of single-particle excitations in correlated electron
systems has been a challenging problem because new phenomena 
are observed there and because simple 
perturbation approaches are not applicable \cite{fulde06}.  
At present, excitations in the systems of infinite dimensions have 
been clarified by use of the dynamical mean field 
theory (DMFT) \cite{georges96}.  
That theory makes use of a momentum-independent self-energy, and determines 
the energy self-consistently by solving an impurity problem with use of
advanced many-body theories.  The DMFT can be traced back to the many-body
coherent potential approximation (CPA) for magnetic alloys 
\cite{hiro77}.  
It is also known
to be equivalent to the dynamical CPA~\cite{kake92}
for describing finite-temperature magnetism \cite{kake04}. 
More recently, we proposed the projection operator method CPA
(PM-CPA) \cite{kake04-3}, 
which is equivalent to the DMFT too.  It is based on the
projection technique applied to the retarded Green function and on an
energy-dependent Liouville operator.

The theories mentioned above are based on a single-site approximation
(SSA) which
neglects the off-diagonal components of the self-energy.  Therefore the
momentum dependence of excitations in real systems is described
incompletely.
In this paper, we propose a nonlocal 
theory of excitations which extends the PM-CPA, and
demonstrate that it describes self-consistently
momentum-dependent excitations with high resolution in the strong
Coulomb interaction regime as well as in the intermediate regime.

We adopt here the Hubbard Hamiltonian $H$ with nearest-neighbor transfer
integral $t$ and intra-atomic Coulomb
interaction $U$, and consider the retarded Green function
\begin{eqnarray}
G_{ij\sigma}(z) = 
\Big( a^{\dagger}_{i \sigma} \  {\Big |} \ \frac{1}{z-L} \,
a^{\dagger}_{j \sigma} \Big) \ .
\label{rg1}
\end{eqnarray}
Here the inner product between two operators $A$ and $B$ is defined as 
$(A|B)=\langle [A^{+},B]_{+} \rangle$ with use of a thermal average 
$\langle \ \ \rangle$ and the anti-commutator $[\ , \ ]_{+}$.
Furthermore $z=\omega+i\delta$ with $\delta$ being an infinitesimal 
positive number,
and $L$ is a Liouville operator defined by $LA=[H,A]_{-}$ for an 
operator $A$.

According to the Dyson equation, the Green function is expressed as
\begin{eqnarray}
G_{ij\sigma}(z) =
[(z - \mbox{\boldmath$H$}_{0} - 
\mbox{\boldmath$\Lambda$}(z))^{-1}]_{ij\sigma}
 \ .
\label{rg2}
\end{eqnarray}
Here $(\mbox{\boldmath$H$}_{0})_{ij\sigma}$ is the Hartree-Fock
Hamiltonian matrix and  
$(\mbox{\boldmath$\Lambda$}(z))_{ij\sigma}$ is the self-energy matrix
defined by 
$\Lambda_{ij\sigma}(z) = U^{2} \overline{G}_{ij\sigma}(z)$.
The reduced memory function $\overline{G}_{ij\sigma}(z)$ is given by
\begin{eqnarray}
\overline{G}_{ij\sigma}(z) = 
\Big( A^{\dagger}_{i \sigma} \  {\Big |} \ \frac{1}{z-\overline{L}} \,
A^{\dagger}_{j \sigma} \Big) \ .
\label{rmem}
\end{eqnarray}
The operator $A^{\dagger}_{i \sigma}$ is defined by 
$A^{\dagger}_{i \sigma}=a^{\dagger}_{i \sigma} \delta n_{i -\sigma}$
with $\delta n_{i -\sigma} = n_{i -\sigma} -  \langle n_{i -\sigma}
\rangle$.
$a^{\dagger}_{i\sigma}$ ($a_{i \sigma}$) is the creation (annihilation)
operator for an electron with spin $\sigma$
on site $i$, and $\langle n_{i\sigma} \rangle$
is the average electron number on site $i$ for spin $\sigma$.
Moreover, $\overline{L}$ is a Liouville operator acting on the
space orthogonal to the space $\{ |a^{\dagger}_{i\sigma}) \}$;
$\overline{L}=QLQ$, $Q=1-P$, and
$P$ is a projection operator defined by
$P= \sum_{i\sigma} \big| a^{\dagger}_{i \sigma} \big) \, 
\big(a^{\dagger}_{i \sigma} \big|$. 

In our nonlocal theory, we introduce an energy-dependent Liouville
operator $\tilde{L}(z)$ for an effective Hamiltonian $\tilde{H}_{0}(z)$
of an off-diagonal medium $\tilde{\Sigma}_{ij\sigma}(z)$, {\it i.e.}, 
$\tilde{H}_{0}(z) = H_{0} + \sum_{ij\sigma} \tilde{\Sigma}_{ij\sigma}(z) \, a^{\dagger}_{i \sigma}a_{j \sigma}$.
Here $H_{0}$ is the Hartree-Fock Hamiltonian. 
The retarded Green function 
$F_{ij\sigma}(z)$ for the Liouvillean $\tilde{L}(z)$ is expressed as 
\begin{eqnarray}
F_{ij\sigma}(z) =
[(z - \mbox{\boldmath$H$}_{0} 
- \tilde{\mbox{\boldmath$\Sigma$}}(z))^{-1}]_{ij\sigma}
 \ ,
\label{fij}
\end{eqnarray}
where $(\tilde{\mbox{\boldmath$\Sigma$}}(z))_{ij\sigma}=\tilde{\Sigma}_{ij\sigma}(z)$.
It should be noted that the Green function $F_{ij\sigma}(z)$ becomes
identical with $G_{ij\sigma}(z)$ when 
\begin{eqnarray}
\tilde{\Sigma}_{ij\sigma} = \Lambda_{ij\sigma}(z) \ .
\label{sceqij}
\end{eqnarray}
Thus, $F_{ij\sigma}(z)$ describes properly many-body
excitations when the above relation is satisfied.

In order to obtain an explicit expression of the self-energy 
$\Lambda_{ij\sigma}(z)$,
we separate the Liouvillean $L$ into $\tilde{L}(z)$ and the
remaining interaction part $L_{\rm I}(z)$, {\it i.e.}, 
$L = \tilde{L}(z) + L_{\rm I}(z)$, and
expand the memory function (\ref{rmem}) by using the incremental
method: 
\begin{eqnarray}
\overline{G}_{ij \sigma}(z) &=&
\overline{G}^{(ij)}_{ij \sigma}(z) + 
\sum_{l \neq i,j} \Delta \overline{G}^{(ijl)}_{ij \sigma}(z)
+ \, \frac{1}{2} \, \sum_{l \neq i,j} \sum_{m \neq i,j,l} \, 
\Delta \overline{G}^{(ijlm)}_{ij \sigma}(z) + \cdots \ .
\label{icrij}
\end{eqnarray}
Here $\Delta\overline{G}^{(ijl)}_{ij\sigma}(z) = 
\overline{G}^{(ijl)}_{ij\sigma}(z) - \overline{G}^{(ij)}_{ij\sigma}(z)$, 
and $\Delta\overline{G}^{(ijlm)}_{ij\sigma}(z) = 
\overline{G}^{(ijlm)}_{ij\sigma}(z) 
- \Delta\overline{G}^{(ijl)}_{ij\sigma}(z) 
- \Delta\overline{G}^{(ijm)}_{ij\sigma}(z) 
- \overline{G}^{(ij)}_{ij\sigma}(z)$.
These terms are all calculated from cluster memory functions defined by 
$\overline{G}^{({\rm c})}_{ij\sigma}(z) =  
\big( A^{\dagger}_{i\sigma} \  {\big |} 
(z-\overline{L}^{({\rm c})}(z))^{-1} \,
A^{\dagger}_{j\sigma} \big)$.
$\overline{L}^{({\rm c})}(z)=QL^{({\rm c})}(z)Q$ and 
$L^{({\rm c})}(z)$ $(c= ij, ijl, ...)$ is the Liouvillean for a cluster
$c$ embedded in the medium with off-diagonal matrix elements 
$\{ \tilde{\Sigma}_{lm\sigma}(z) \}$.
Note that a ``cluster'' $c=(ij)$, for example, does not mean that sites
$(i,j)$ are nearest neighbors.  Instead they may be far apart.

When we truncate the expansion at a certain stage in the
series of Eq. (\ref{icrij}), the self-energy 
$\Lambda_{ij\sigma}(z)=U^{2}\overline{G}_{ij\sigma}(z)$ depends on the
medium $\tilde{\Sigma}_{ij\sigma}(z)$.  We then determine the medium
self-consistently from condition (\ref{sceqij}).  
Note that the present theory reduces to the PM-CPA ({\it i.e.}, the
DMFT) in
infinite dimensions.

In the above self-consistent theory, one needs explicit expressions for
the cluster memory functions, for which we apply the renormalized 
perturbation theory (RPT) \cite{kake04-2}.
\begin{eqnarray}
\overline{G}^{({\rm c})}_{ij\sigma}(z) = 
\Big[ \overline{\mbox{\boldmath$G$}}^{({\rm c})}_{0}(z) \cdot 
(1-\overline{\mbox{\boldmath$L$}}^{({\rm c})}_{I} \cdot 
\overline{\mbox{\boldmath$G$}}^{({\rm c})}_{0}(z))^{-1} \Big]_{ij\sigma} \ .
\label{memcij}
\end{eqnarray}
Here $(\overline{\mbox{\boldmath$L$}}^{({\rm c})}_{I})_{i\sigma
j\sigma^{\prime}} = U (1- 2 \langle n_{i -\sigma}
\rangle)/\chi_{i\sigma}$, and
$\chi_{i\sigma} = \langle n_{i-\sigma} \rangle 
(1 - \langle n_{i-\sigma} \rangle)$. 
\begin{figure}[t]
\begin{center}
\includegraphics[width=30pc]{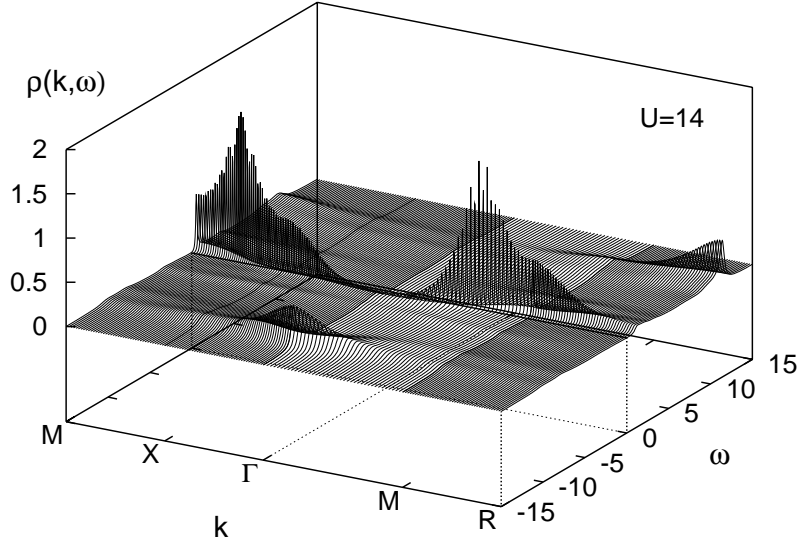}%
\end{center}
\vspace*{-10mm}
\caption{\label{label}
Momentum-dependent spectra along high-symmetry lines at $U=14$.
The Fermi level is indicated by a bold line.
}
\end{figure}
%
%
In the simplest approximation (RPT-0), 
$(\overline{\mbox{\boldmath$G$}}^{({\rm c})}_{0})_{ij\sigma}(z)$ is given by
\begin{eqnarray}
(\overline{\mbox{\boldmath$G$}}^{({\rm c})}_{0})_{ij\sigma}(z) = 
A_{ij\sigma}
\int \frac{\displaystyle
d\epsilon d\epsilon^{\prime} d\epsilon^{\prime\prime} 
\rho^{({\rm c})}_{ij\sigma}(\epsilon)
\rho^{({\rm c})}_{ij-\sigma}(\epsilon^{\prime})
\rho^{({\rm c})}_{ji-\sigma}(\epsilon^{\prime\prime})
\chi(\epsilon, \epsilon^{\prime},
\epsilon^{\prime\prime})
}
{\displaystyle 
z - \epsilon - \epsilon^{\prime} + \epsilon^{\prime\prime}}
\ .
\label{rpt00}
\end{eqnarray}
Here 
$A_{ij\sigma} = \chi_{i\sigma} /
\langle n_{i -\sigma} \rangle_{\rm c} 
(1 - \langle n_{i -\sigma} \rangle_{\rm c}) \delta_{ij} + 1 - \delta_{ij}$,
$\chi(\epsilon, \epsilon^{\prime}, \epsilon^{\prime\prime}) =
(1-f(\epsilon))(1-f(\epsilon^{\prime})) f(\epsilon^{\prime\prime})
+ f(\epsilon)f(\epsilon^{\prime})
(1-f(\epsilon^{\prime\prime}))$, 
$f(\epsilon)$ is the Fermi distribution function, 
$\langle n_{i\sigma} \rangle_{\rm c}$ is the electron number for a
cavity state defined by $\langle n_{i\sigma} \rangle_{\rm c} = \int d\epsilon 
\rho^{({\rm c})}_{ii\sigma}(\epsilon) f(\epsilon)$, and 
$\rho^{({\rm c})}_{ij\sigma}(\epsilon) = - \pi^{-1} {\rm Im} [(
\mbox{\boldmath$F$}_{\rm c}(z)^{-1}
+ \tilde{\mbox{\boldmath$\Sigma$}}^{({\rm c})}(z))^{-1}]_{ij\sigma}$.
The coherent cluster Green function 
$(\mbox{\boldmath$F$}_{\rm c}(z))_{ij\sigma}=F_{ij\sigma}(z)$ is given by 
Eq. (\ref{fij}), and 
$(\tilde{\mbox{\boldmath$\Sigma$}}^{({\rm c})}(z))_{ij\sigma}=
\tilde{\Sigma}_{ij\sigma}(z)$ for sites $(i,j)$ belonging to the cluster $c$.

Finally the momentum dependent excitation spectra are calculated from
the Green function
\begin{eqnarray}
G_{k\sigma}(z) = \frac{1}{z - \epsilon_{k\sigma} - 
\Lambda_{k\sigma}(z)} \ .
\label{gk}
\end{eqnarray}
Here $\epsilon_{k\sigma}$ is the Hartree-Fock one-electron energy 
eigen value, 
and the momentum-dependent self-energy is calculated via Fourier
transform of the off-diagonal self-energy as
$\Lambda_{k\sigma}(z) = \sum_{j} \Lambda_{j0\sigma}(z) 
\exp (i\mbox{\boldmath$k$}\cdot\mbox{\boldmath$R$}_{j})$.

We have performed numerical calculations for the single-particle
excitation spectra on a simple cubic lattice at half-filling.  In the
nonlocal self-energy calculations, we have taken all two-site pairs
up to 10-th nearest neighbors,
and neglected the clusters beyond the 2 sites.  Figure 1 shows an
example of the momentum-dependent
spectra along high-symmetry lines for $U=14$ in unit of $|t|=1$.  
We find quasiparticle states near the Fermi level, which form a narrow
band with averaged quasiparticle weight $Z=0.21$.
On the other hand, the lower (upper) Hubbard band is enhanced around 
the $\Gamma$ (R) point.  It should be noted that the
energy splitting between the main peaks of the lower and upper Hubbard
bands is about 20, 
being larger than $U=14$ as seen in Fig. 2.  This seems to be 
explained by strong antiferromagnetic (AF) correlations.  In fact, 
in the strong AF correlation limit the energy to remove (add) an 
electron is expected to be $\epsilon_{0}-zJ$ ($\epsilon_{0}+U+zJ$) at
half filling.  Here $\epsilon_{0}$ is the atomic level and 
$J$ is the super exchange interaction constant  
($J=4|t|^{2}/U$).  Then the splitting is given by
$U+2zJ$ instead of $U$.  The former yields 17.4 for $U=14$, which is 
comparable to the calculated splitting 20.  We also find sub-peaks
around $|\omega|=8$ in Figs. 1 and 2.  
These peaks are interpreted as the lower and upper
Hubbard peaks in the absence of long-range AF correlations because the
splitting is close to $U$.  Furthermore, we find weak excitations at
$|\omega| \approx 3$, which might correspond to a ``shadow band'' due to
strong AF correlations as found in QMC calculations \cite{grober00}.
We also point out that in the nonlocal theory, the density of 
states at the Fermi level $\rho(0)$ does not agree with the one in 
the SSA as seen in Fig. 2.  

\begin{figure}[t]
\includegraphics[width=18pc]{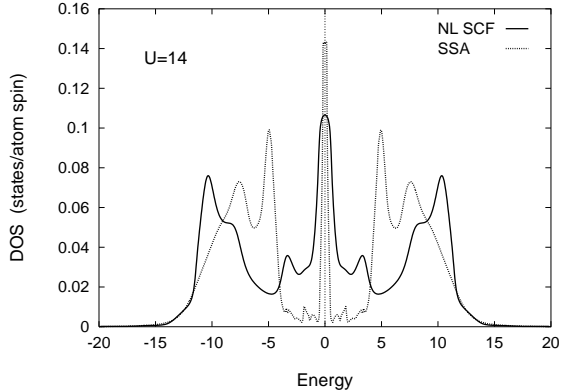}\hspace{2pc}%
\begin{minipage}[b]{18pc}\caption{\label{label}
Densities of states in the present theory (solid curve) and the SSA
 (dotted curve).
}
\end{minipage}
\end{figure}

We have calculated the nonlocal excitation spectra with increasing Coulomb
interaction.  Self-consistent solutions are found up to $U \approx
21$.  In this range, we find that the quasiparticle weights $Z$ are larger
than those of the SSA but never vanish.
We suggests that the critical Coulomb interaction is 
$U_{c2}(Z=0) \gsim 30$.  

In summary, we have presented a nonlocal theory for excitation 
spectra by introducing a fully off-diagonal effective medium
$\Sigma_{ij\sigma}(z)$ to the projection operator formalism.  The theory
describes self-consistently the long-range nonlocal excitations in
momentum space with high resolution.  We verified from the numerical
calculations 
that the present scheme works from the weak to strong Coulomb
interaction regime, and found the shadow bands as well as
the Hubbard sub-bands due to strong AF correlations.
\vspace{2mm}

This work was supported by the Grant-in-Aid for Scientific Research (19540408).


\section*{References}
\vspace{1mm}
\begin{thebibliography}{9}
\bibitem{fulde06} Fulde P, Thalmeier P, Zwicknagl G 2006, Solid State
	Phys. {\bf 60} 1
\bibitem{georges96} Georges A, Kotliar G, Krauth W, Rosenberg M J 1996
	Rev. Mod. Phys. {\bf 68} 13
\bibitem{hiro77} Hirooka S and Shimizu M 1977 J. Phys. Soc. Jpn. {\bf
	43} 70
\bibitem{kake92} Kakehashi Y 1992 Phys. Rev. B {\bf 45} 7196;
	2002 Phys. Rev. B {\bf 65} 184420
\bibitem{kake04} Kakehashi Y 2004 Adv. in Phys. {\bf 53} 498
\bibitem{kake04-3}
Kakehashi Y and Fulde P 2004 Phys. Rev. B{\bf 69} 045101
\bibitem{kake04-2}
Kakehashi Y and Fulde P 2004 Phys. Rev. B{\bf 70} 195102
\bibitem{grober00}
Gr\"ober C, Eder R, and Hanke W 2000 Phys. Rev. B{\bf 62} 4336
\end{thebibliography}
\end{document}